\documentclass[12pt,preprint]{aastex}

\newcommand{\Ms}{M$\sun$}

\shorttitle{Molecules  in massive primordial supernovae}
\shortauthors{Cherchneff}

\begin{document}


\title{Primordial massive supernovae as the first molecular factories in the early universe}


\author{Isabelle  Cherchneff \& Simon Lilly\altaffilmark{1,2}}




\altaffiltext{1}{Intitut f{\"u}r Astronomie, ETH H{\"o}nggerberg, 8093 Z{\"u}rich, Switzerland.}
\altaffiltext{2}{isabelle.cherchneff@phys.ethz.ch, simon.lilly@phys.ethz.ch}


\begin{abstract}
We study the ejecta chemistry of a zero-metallicity progenitor, massive, supernova using a novel approach based on chemical kinetics. Species considered span the range of simple, di-atomic molecules such as CO or SiO to more complex species involved in dust nucleation processes. We describe their formation from the gas phase including all possible relevant chemical processes and apply it to the ejecta of a primordial 170  \Ms~supernova. Two ejecta cases are explored:  full mixing of the heavy elements, and a stratified ejecta reflecting the progenitor nucleosynthesis. Penetration of hydrogen from the progenitor envelope is considered. 
We show that molecules form very efficiently in the ejecta of primordial supernovae whatever the level of mixing and account for 13 to 34\% of the total progenitor mass, equivalent to 21 to 57 \Ms~of the ejecta material in molecular form. The chemical nature of molecules depends on mixing of heavy elements and hydrogen in the ejecta. Species produced include O$_2$, CO, CO$_2$, SiS, SO, SiO and H$_2$. Consequently, molecules can be used as observational tracers of supernova mixing after explosion. We conclude that primordial massive supernovae are the first molecule providers to the early universe. 
\end{abstract}




\section{Introduction}

Large amounts of dust have been conjectured to explain the reddening of background quasars and damped Ly$\alpha$ systems in the early universe (Pettini et al. 1994, Pei \& Fall 1995). Possible dust makers could be primordial, very massive stars exploding as supernovae, for the time scales at redshifts $\ge$ 6 imply short stellar evolution times, thus excluding low-mass, evolved stars. Dust formation in such massive objects has been studied (Nosawa et al. 2003, Schneider et al. 2004) using a classical nucleation theory and excluding details on nucleation processes of dust clusters from the gas phase. However, it is now well accepted that dust forms in other evolved stellar environments under non-equilibirum conditions close to those encountered in laboratory dust condensation experiments where chemical kinetics commands dust nucleation from the gas phase (Donn \& Nuth 1985, Cherchneff et al. 1992). In supernovae like in other environments, the nucleation will take place via the formation of a molecular phase in the ejecta. Detection of CO, SiO and H$_3^+$ in SN1987A, a core-collapse supernova in the Large Magellanic Cloud (Spyromilio et al. 1988, Roche et al. 1991, Miller et al. 1992), triggered a few theoretical studies on modeling CO and SiO observations (Petuchowski et al. 1989, Lepp et al. 1990, Liu \& Dalgarno 1994, Gearhart et al. 1999). Those models often used incomplete chemical networks at steady state. No complete chemical description of a primordial, massive supernova ejecta has been so far attempted. We present here the first results of such an endeavor and consider as a surrogate a primordial massive supernova (hereafter PMSN) with zero-metallicity 170  \Ms~progenitor. We study the formation of chemical species, some of which being dust precursors, in its ejecta. We show that chemistry is usually not at steady state and fosters the formation of complex molecules. Up to one third of the ejecta is found to be in molecular form depending on the level of mixing after explosion.

\section{Ejecta physical and chemical model}

In the absence of observational constraints on PMSNe, we base our study on the 170 \Ms~PISN theoretical explosion model developed by Nozawa et al. (2003). Gas temperature T is determined mainly by the explosion energy ($=$ 2$\times 10^{52}$ ergs) and by solving the radiative transfer equation taking into account the energy deposition by radioactive elements. We choose the T profile for  their 170 \Ms~PISN unmixed case to describe our ejecta temperature variation with time t and fit it with the expression ${\rm T(M_r,t)= T_0(M_r,t_0) \times (t/t_0)^{3(1-\gamma)}}$,
where M$_r$ is the mass coordinate, T$_0$ is the gas temperature at t$_0$, and $\gamma$ is a parameter equals to 1.55 for this specific fit. The ejecta expansion is homologous very early on and the gas density varies with time according to ${\rm n(M_r,t) = n_0(M_r, t_0)\times (t/t_0)^{-3}}$,
where n$_0$ is the gas number density at t$_0$. 
We are interested in studying chemistry from t$_0$ = 100 days to t=1000 days for  we expect molecules and dust to form at gas temperatures smaller than 4000K. High temperatures thermally destroy molecules at early times whereas the gas parameters have too low values to foster efficient molecular formation at late times. In SN1987A ejecta, which is characterized by lower gas temperatures than those found in the present PMSN model, CO, SiO and dust are detected as early as 110, 160, and 450 days, respectively (Catchpole \& Glass 1987, Wooden et al. 1993). The ejecta velocity is kept constant at 2000 Km s$^{-1}$, a value similar to that derived by Schneider et al. (2004) for a 170 \Ms~progenitor. Values for the ejecta physical parameters are summarized in Table 2. 

SN1987A light curve between 100 and 1000 days after explosion is dominated by the radioactive decay energy of $^{56}$Co and is well reproduced if 0.075 \Ms~of $^{56}$Co is assumed to be produced over mass cut. $\gamma$-rays produced from $^{56}$Co radioactive decay are Compton-scattered, producing fast, energetic electrons, which are one of the dominant destruction processes to molecules. We assume that similar radioactivity-induced processes take place in the ejecta of our PMSN, but choose a  $^{56}$Co mass scaled on the  $^{56}$Ni mass over mass cut according to the model of Umeda \& Nomoto (2002)~(M($^{56}$Co) = 3.52 \Ms). The energy deposition by thermalized $^{56}$Co $\gamma$-rays is proportional to M($^{56}$Co) and the destruction rate by Compton electrons for species {\it i} in s$^{-1}$ is given by (Woosley et al. 1989, Liu \& Dalgarno 1995)
\begin{equation}
{\rm  k_{fast~e^{-}}({\it i}) =( 3.26\times 10^{-2}/W{\it_i}) \exp(-t/\tau_{56}) \times (1-\exp[-\tau_0(t/t_0)^{-2}])}.
\end{equation}
where $\tau_{56}$, $^{56}$Co decay e-folding time, equals 111.26 days, $\tau_0 =31.1$ at t$_0$, and W{\rm $_i$} is the mean energy per ion-pair for species {\it  i} in eV. We do not consider ultraviolet radiation coming from ${\gamma}$/X-rays-induced excited atoms as molecules are likely to experience self- or mutual-shielding in the ejecta dense clumps (Liu \& Dalgarno 1996). The early emergence of $\gamma$-rays from $^{56}$Co decay in SN1987A spectra argues for strong mixing by Rayleigh-Taylor instabilities after explosion (Pinto \& Woosley 1988, Kumagai et al. 1988). Such instabilities are to be expected in PMSNe where the explosion energy is far greater, resulting in macroscopic and/or microscopic mixing. We thus assume two extreme mixing cases as in Nozawa et al. (2003): a fully microscopically-mixed ejecta in which some level of H penetration from the progenitor envelope occurs, and a stratified ejecta retaining the onion-like structure due to stellar nucleosynthesis. Chemical composition of the fully-mixed case is that of Umeda \& Nomoto (2002) for a 170 \Ms~PISN whereas the stratified composition of Nozawa et al. (2003) is used for the unmixed case. We subdivide their unmixed, 85 \Ms~helium core into 4 regions: a 20 \Ms~Si/Fe-rich zone, a 50 \Ms~O/Si/Mg-rich layer, a 10 \Ms~O/C/Mg-rich zone and finally a 5 \Ms~H/He/C-rich layer. 

The chemical network encompasses all chemical processes appropriate to our ejecta gas parameters. They include tri-molecular reactions efficient in high density media, bi-molecular processes like neutral-neutral reactions with/without activation energy, and ion-molecule reactions (formation/destruction and charge exchange reactions). Radiative associations (hereafter RA) are also considered. In total, the system comprises 79 species listed in Table 1 and between 400 to 500 reactions, depending on the ejecta region under study. We decouple the dynamics and the chemistry and integrate 79 stiff, non-linear, coupled, ordinary differential equations describing the continuity equation for the chemical species. The chemical network is applied to a physical model of SN1987A ejecta to test its relevance to molecular formation. The unmixed chemical composition of Woosley (1988 - see Fig 4) is used and we divide the 6 \Ms-He core C-rich region into two zones: zone 1 extending from 3.5 to 4 \Ms~and zone 2, from 4 to 6 \Ms. Figure 1 displays our modeled CO mass compared to CO masses derived from different fits of the CO first-overtone band in the infrared. The agreement is satisfactory in view of the different assumptions and parameters used to fit the measured spectra (i.e., LTE or non-LTE assumption for level populations (Spyromilio et al. 1988, Liu et al. 1992), cool clump model including CO cooling (Liu \& Dalgarno 1995)). Similar agreements are found for SiO and H$_3^+$ and are presented in Cherchneff (2008).  

\section{Results and conclusions}

Molecular abundances with respect to the total gas number density are displayed in Figure 2 for the fully-mixed case. Here, we consider two sub-cases: the extreme case where the entire H envelope is microscopically mixed to the He core (hereafter referred as H-rich case), and a case for which only 1\% of the H envelope mixes with the He-core (i.e. H-poor case). In both cases several molecules do form but differences exist among the two situations. 

For the H-rich case, the dominant ejected species at 1000 days are H$_2$, O$_2$, SO, CO$_2$, and N$_2$. Their corresponding masses are respectively 19.3 \Ms, 15.3 \Ms, 13.0 \Ms, 8.4 \Ms, and 1.0 \Ms, thereby implying a total molecular mass of $\sim$ 57 \Ms~equivalent to 34\% of the PMSN progenitor mass. SiO is abundant up to t = 550 days, but is rapidly depleted due to silica/quartz precursor formation. Inspection of Figure 2 shows that, chemically speaking, the steady state assumption does not hold for molecular formation and destruction at t $\le$ 1000 days. At early times, the dominant chemical processes at play are neutral-neutral and RA reactions. The OH radical is a key species to molecular formation. Destruction occurs mainly via He$^+$ attack. For t$\ge$ 600 days, neutral-neutral processes without activation barrier and ion-molecule reactions are dominant. As for dust precursors, (SiO$_2$)$_2$, a ring precursor to silica/quartz nucleation, forms in large amount --  33.3 \Ms, as early as 500 days after explosion, followed by AlO, the gas phase precusor to corumdum (Al$_2$O$_3$) -- 0.04~\Ms~at 530 days, and finally (FeO)$_2$ -- 0.001~\Ms~at 750 days. For the H-poor case, the dominant molecular species are SiO - 15.8~\Ms, CO - 4.6~\Ms~and CO$_2$ - 1~\Ms, resulting in a total molecular content of 21.5~\Ms, or 12.6\% of the progenitor mass. When H is absent, the formation chemistry is powered at early times by RA reactions as opposed to neutral-neutral reactions involving OH. The RA forming SiO has a reaction rate 50 times greater than that for O$_2$ formation, resulting in large amounts of SiO compared to O$_2$. Molecular  formation by  neutral-neutral processes without activation barrier is postponed to later times, as seen in Figure 2. A smaller molecular content is thus produced due to lower gas temperature and density. As for dust, (SiO$_2$)$_2$ again is the dominant precursor to form  at level up to 10.7~\Ms, while (MgO)$_2$, AlO and (FeO)$_2$ contents are negligible. 

The above dust nucleation sequences are different from those of existing classical nucleation studies which rely on condensation temperature analysis for solids (Todini \& Ferrrara 2001, Nozawa et al. 2003, Schneider at al. 2004). Such studies predict the condensation of corundum, forsterite (Mg$_2$SiO$_4$), quartz (SiO$_2$), amorphous carbon and magnetite (Fe$_3$O$_4$). The discrepancy resides in the approach used: in the present model, molecules form from the gas simultaneously to dust precursors, thereby depleting some of the available elements from the gas phase. Dust precursor formation is then commanded by chemical kinetics at play in an altered gas phase whose chemical composition differs drastically from the initial elemental composition. The absence of carbon dust precursors in our fully-mixed case illustrates this point. Indeed, carbon is locked up in CO and CO$_2$ despite the inclusion of Compton electron destruction reactions (Clayton et al. 1999), and is not available for further build-up of C-rich molecules and amorphous carbon precursors. Hence chemistry is acting as a bottleneck to dust nucleation. If we assume that all gas-phase precursors formed in our fully-mixed ejecta are included into dust during condensation, we get an upper limit for freshly formed dust of roughly 33.4~\Ms~in the H-rich case and 10.7~\Ms~in the H-poor case, equivalent to $\sim 19\%$ and 6\% of the PMSN progenitor mass, respectively. However, such an assumption overestimates the dust content as dust condensation efficiencies in pyrolysis or flame experiments in the laboratory are usually less than one, with a residual population of dust precursors in the gas phase (J{\"a}ger et al. 2006). Furthermore, our H-rich case is somehow extreme and the total H envelope mixing with the He core unlikely, though H mixing was observed in SN1987A and should occur in PISNe as well. Nozawa et al. (2003) and Schneider et al. (2004) find respectively  $\sim$ 33 \Ms~and $\sim$ 39 \Ms~of dust for the fully-mixed He core of a 170 \Ms~progenitor mass PISN and similar ejecta thermodynamics than that in the present study. These results are to be compared to the 10.7 \Ms~upper limit derived for our H-poor case. Our two mixed cases highlight trends in dust formation scenarios and content in PMSNe and our limits for freshly formed dust are a few \% of the progenitor mass only. This is certainly lower than existing predicted dust amounts.

For the unmixed ejecta, molecules and their derived mass are listed in Table 2. Again, molecules do form efficiently and account for $\sim$ 42~\Ms, equivalent to 25\% of the PMSN progenitor mass. Their chemical nature now traces their location in the He core and the amount of H mixing from the progenitor envelope. In Zone 1, devoid of oxygen, SiS forms from reaction between atomic Si and S$_2$ and traces a Si/S-rich gas. In Zone 4, we allowed for 18 \% of total mass hydrogen penetration in the C/He--rich layer as in Nosawa et al. (2003), with subsequent formation of H$_2$, C$_2$H$_2$, and CO. Carbon monoxide is preferably formed over CO$_2$ in this layer as CO$_2$ is rapidly destroyed by reactions with ions like C$^+$ and He$^+$ to re-form CO. Thus, CO presence in the ejecta indicates a C/O ratio greater than 1, a large free carbon atom content, and some mixing with He and H. This result is supported by the simultaneous detection of CO and amorphous carbon dust in SN1987A, implying carbon-rich inhomogeneities in the ejecta. In Zone 2, CO$_2$ dominates over CO, because of the rapid CO conversion to CO$_2$ via reaction with O$_2$, which is very abundant in those regions where the C/O ratio is less than 1. CO$_2$ is thus a tracer of oxygen-rich ejecta regions whereas the molecular composition of Zone 4 is typical of a C/H/He-rich environment. Similar chemical processes and species are encountered in the inner shocked winds of O-rich and C-rich, evolved low-mass stars, where the gas parameters resemble those of SN ejecta (Cherchneff 2006).  As for dust molecular precursors, we find that Zone 1 produces $\sim$ 1.6~\Ms~of (Si)$_4$ and $\sim$ 1.3~\Ms~of (FeS)$_2$, while Zone 2 produces 3~\Ms~of (SiO$_2$)$_3$. No dust precursors are formed in Zone 3, and Zone 4 produces 0.5~\Ms~of C$_3$. Again, grain types and nucleation sequences are different from the study of Nosawa et al. (2003). An upper limit to dust production in the unmixed case is $\sim$ 6.5 \Ms~or $\sim$ 3.8$\%$ of the total PMSN progenitor mass, which is much less than the value of $\sim$ 18 \% derived by Nozawa et al. for their 170 \Ms~PISN unmixed case. This points again to the crucial role of chemical kinetics as bottleneck to dust formation. Chemically speaking, we notice that Zone 4 forms carbon dust precursors via pure carbon chains like C$_3$ despite the large quantities of C$_2$H$_2$ available and the aromatic formation reactions included in our chemical network. Aromatic chemical pathways to carbon dust are inhibited even in the presence of hydrogen because of the high temperatures encountered in the ejecta up to 600 days after explosion, and the competitive RA pathways to carbon chains combined to large amounts of free C atoms. This result may be specific to supernova ejecta and is supported by the non-detection of polycyclic aromatic hydrocarbon infrared emission lines in SN1987A spectra (Wooden et al. 1993). 

We conclude that Pop III, massive supernovae are efficient molecule providers to the early universe, with 13\%  to 34$\%$ of their ejecta in molecular form, corresponding to 22 \Ms~to 57~\Ms~of molecular material released to the local, pristine gas after explosion. Formed species depend on mixing in the ejecta and include O$_2$, CO$_2$, SO, CO, SiS, OH, C$_2$H$_2$ and H$_2$. H mixing boosts molecular formation at early times via neutral-neutral processes, resulting in a large molecular component in the ejecta. In a more general context, molecules could be used as observational tracers of mixing in nearby core-collapse supernovae. The substantial amounts of molecular material imply some impact on the local gas cooling if cooling time scales are comparable to those for the ejecta adiabatic expansion. It was recently shown that some dust grains could survive the passage of the PMSN reverse shock some 10$^4$ years after explosion (Nozawa et al. 2007, Bianchi \& Schneider 2007). The survival of molecules in cool, dense, inhomogeneities passing the reverse shock must then be studied. At later times, it is conjectured that Pop.~II.5 stars can form in the PMSN dense shell if other cooling than that of H$_2$ is provided (MacKey et al. 2003, Salvaterra et al. 2004). If surviving the reverse shock, our predicted  molecules could provide part of the necessary cooling and their subsequent impact on second generation star formation needs further investigation. Finally, under present chemical kinetic conditions, our results show that PMSNe do form dust efficiently as the amounts of dust precursors produced are significant. However, the present study points to lower dust contents formed in PMSNe than those predicted by existing, classical nucleation studies.
\acknowledgments

We are grateful to the anonymous referee for useful suggestions on how to improve the manuscript. IC acknowledges support from a Maria-Heim-V{\"o}gtlin Fellowship from the Swiss National Science Fundation.

\clearpage


\begin{deluxetable}{llllllllllll}
\tabletypesize{\scriptsize}
\tablewidth{0pt}
\tablecaption{Chemical species included in the 170 \Ms~PMSN ejecta model}             
\startdata  
\tableline\tableline
Atoms& H & He & O & C & Si &S  & Mg &Fe & Al & &  \\
\tableline
 Diatomic  species& H$_2$ & OH & O$_2$ & CO & SiO & SO&NO  & MgO &FeO & AlO  &C$_2$  \\
 & CS&  CN & SiH  &SiC & Si$_2$ & SiS & SiN  &SH & N$_2$ & NH & MgS  \\
 & Fe$_2$  & & & & & & & & & & \\
\tableline
Tri-atomic species& H$_2$O & H$_2$S & HCN & CH$_2$ & C$_2$H &HCO &C$_3$ & CO$_2$ & OCS & OCN  & SiC$_2$  \\
 & Si$_3$&SiO$_2$&SO$_2$& NO$_2$&Fe$_3$& & & & & \\
\tableline
4-atom species &  C$_2$H$_2$ & Si$_2$O$_2$ & Mg$_2$O$_2$ & Mg$_2$S$_2$ &Fe$_2$O$_2$ & Fe$_2$S$_2$ &H$_2$CC &Fe$_4$&Si$_4$& &  \\
\tableline
$\ge 5$-atom species& Si$_2$O$_4$ &Si$_3$O$_6$&C$_3$H$_3$&C$_4$H$_4$& & & & & &  \\  
\tableline
Ions & H$^+$ & H$^-$ & He$^+$ & O$^+$&Si$^+$&S$^+$&Mg$^+$ &Fe$^+$&Al$^+$ & H$_2^+$  &H$_3^+$  \\
 & HeH$^+$& C$_2^+$& CO$^+$ &SiO$^+$ &SO$^+$ & H$_2$O$^+$ & HCO$^+$ & & &   \\                            
\enddata
\end{deluxetable}

\clearpage


\begin{deluxetable}{ccc}
\tabletypesize{\scriptsize}
\tablecaption{170 \Ms~PMSN ejecta parameters versus time after explosion.}
 \tablewidth{0pt}
 \tablehead{
\colhead{Time (days) } & \colhead{Temperature (K)} & \colhead{Number density (cm$^{-3}$)}  
}
\startdata
100 & 21000 & 6.50 $\times 10^{11}$ \\
200  & 6640 &8.00 $\times 10^{10}$  \\
400 &  2130 & 1.02 $\times 10^{10}$ \\ 
600  & 1090 & 3.01 $\times 10^{9}$ \\
800 &677 & 1.26 $\times 10^{9}$\\
1000  &  470 &6.50 $\times 10^{8}$   
 \enddata

\end{deluxetable}
\clearpage
\begin{deluxetable}{lcc}
\tabletypesize{\scriptsize}
\tablecaption{Molecules excluding dust precursors in the unmixed ejecta of a 170 \Ms~PMSN}
\tablewidth{0pt}
\tablehead{
\colhead{ } & \colhead{Species} & \colhead{Mass (\Ms)}  
}
\startdata
Zone 1 & SiS &  11.60  \\
(20 \Ms) & S$_2$ &  1.48$\times 10^{-4}$ \\ 
\tableline
  Zone 2 & O$_2$ & 20.53\\
(50 \Ms)   & SO & 1.05 \\
 & CO$_2$ & 1.74$\times 10^{-3} $ \\
 \tableline
   Zone 3 &  CO & 4.28\\
(10 \Ms)     & CO$_2$ & 0.43 \\
 &SiO&0.11 \\
\tableline
   Zone 4 & CO & 3.45\\
(5 \Ms)       & C$_2$H$_2$ & 0.43\\
         & H$_2$ & 0.15  \\
\tableline
Total & &  {\bf 42.03} 
\enddata
\end{deluxetable}

\clearpage
\begin{figure}
\epsscale{1.2}
\plotone{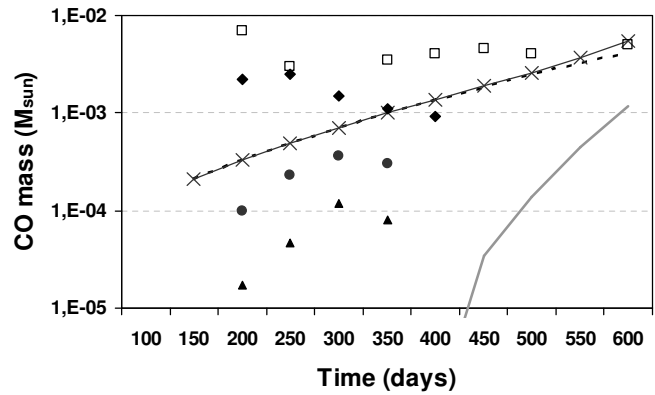}
\caption{Predicted CO mass in SN1987A as a function of time compared to CO mass derived from observations. Dotted line:  zone 1 (see text for explanation), continuous grey line: zone 2, continuous grey-crossed line: C-rich region of our unmixed model (zone 1+ zone 2), filled triangle: data from Spyromilio et al. 1989, filled circle: data from Liu et al. 1992 - LTE assumption, filled diamond: data from Liu et al. 1992 - non-LTE assumption, open square: data from Liu \& Dalgarno 1995 - low T clump. }
\end{figure}
\clearpage
\begin{figure}
\epsscale{1.2}
\plottwo{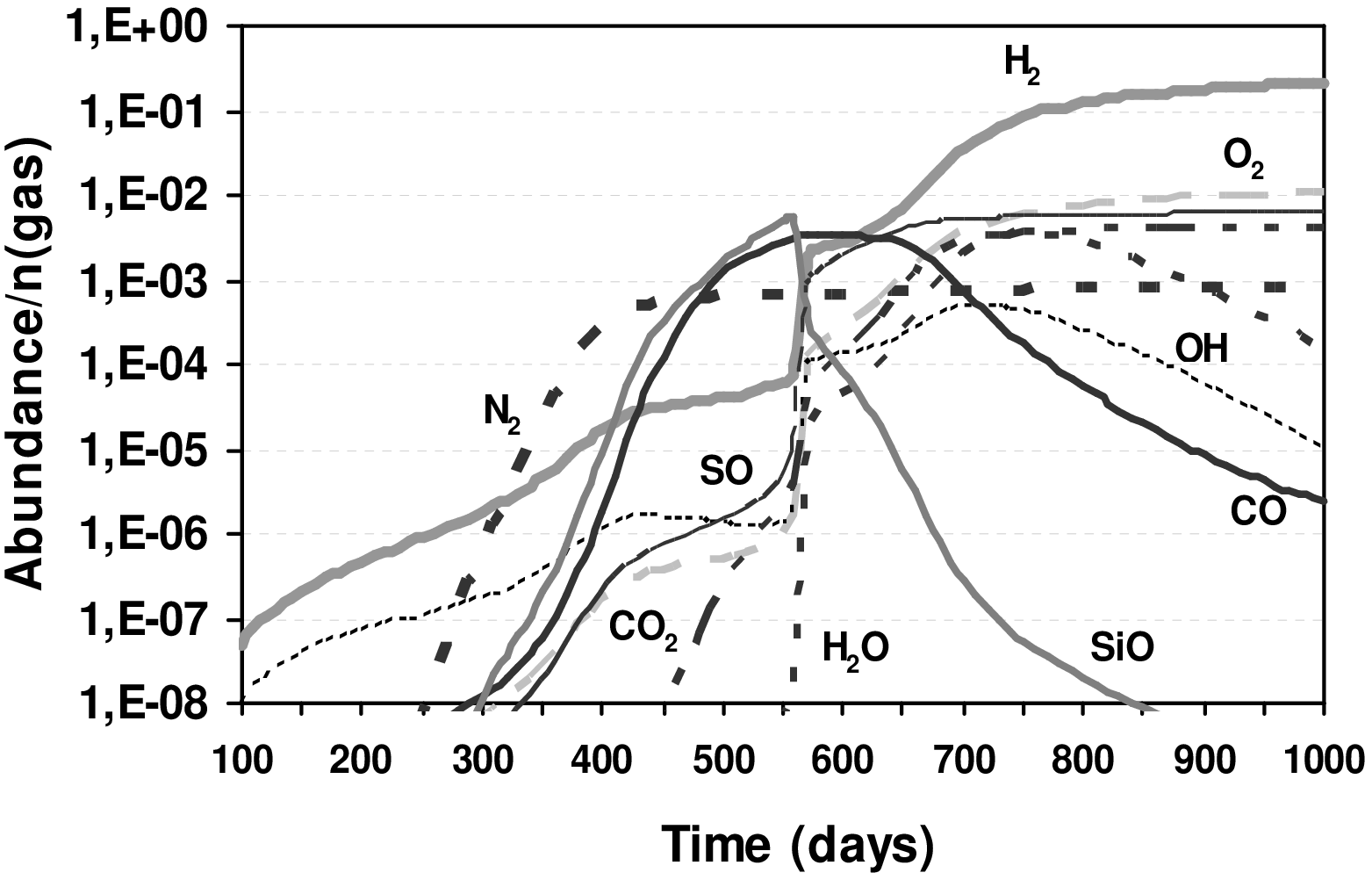}{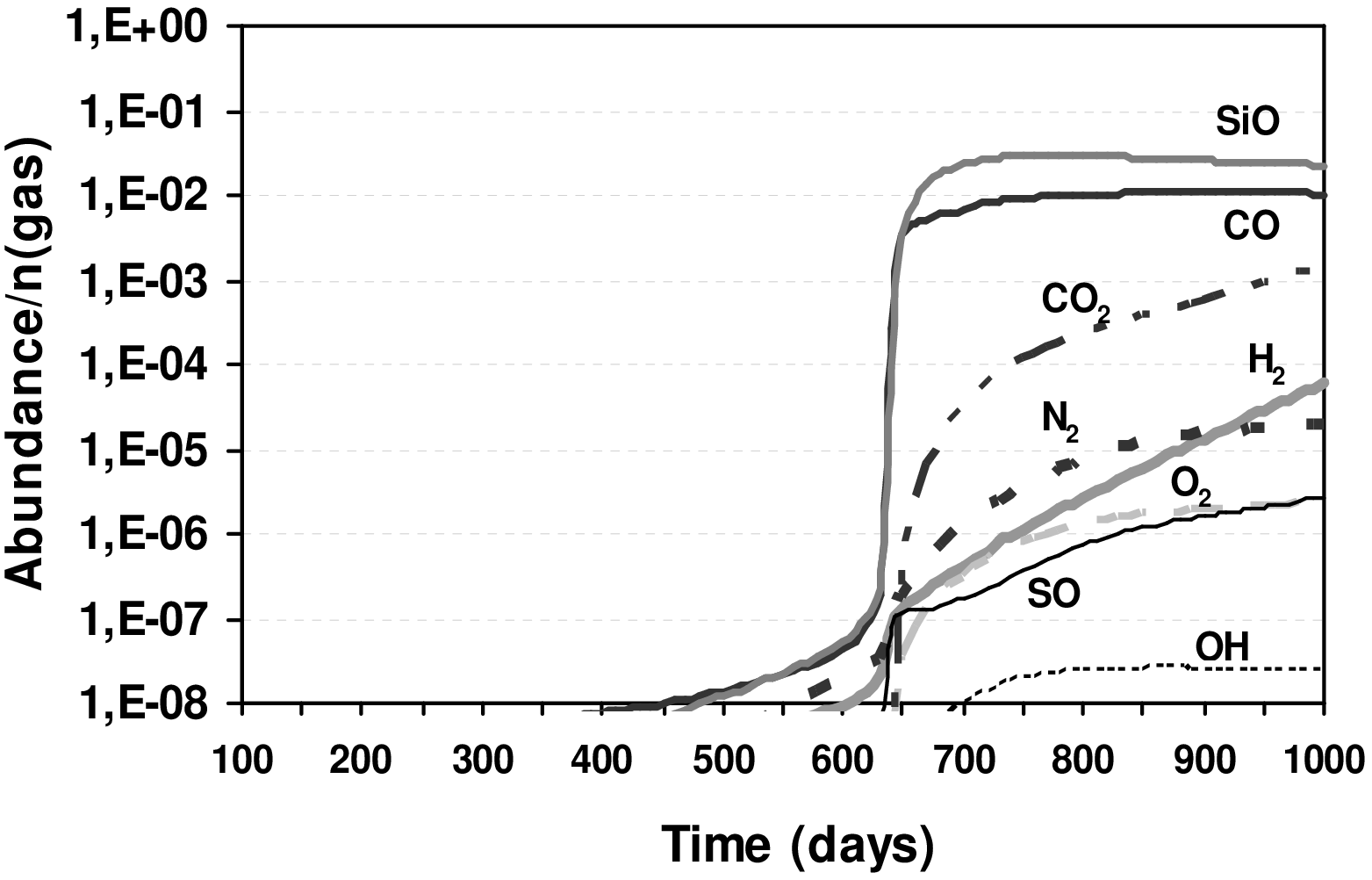}
\caption{Molecular abundances with respect to total gas number density as a function of time after explosion for the fully-mixed 170 \Ms~PMSN case of Umeda \& Nomoto (2002). Left) The total H envelope is fully mixed to the He core. Right) Only 1\% of the H envelope is mixed to the He core.}
\end{figure}

\clearpage

\end{document}